\documentclass{mem}
\usepackage{natbib}\usepackage{txfonts}\usepackage{balance}
\usepackage{graphicx}
\usepackage[a4paper,breaklinks,dvipdfm]{hyperref}
\usepackage[rightcaption]{sidecap}
\idline{1}{1}
\begin{document}

\title{
The effect of intermediate mass close binaries on the chemical evolution of Globular Clusters
}

   \subtitle{}

\author{
Nicki Mennekens, Dany Vanbeveren \and Jean-Pierre De Greve
          }

\institute{
Astrophysical Institute, Vrije Universiteit Brussel, Pleinlaan 2, 1050 Brussels, Belgium\\
\email{nmenneke@vub.ac.be}
}

\authorrunning{Mennekens et al.}

\titlerunning{The effect of IMCBs on the chemical evolution of GCs}

\abstract{
The chemical processes during the Asymptotic Giant Branch (AGB) evolution of intermediate mass single stars predict most of the observations of the different populations in Globular Clusters although some important issues still need to be further clarified. In particular, to reproduce the observed anticorrelations of Na-O and Al-Mg, chemically enriched gas lost during the AGB phase of intermediate mass single stars must be mixed with matter with a pristine chemical composition. The source of this matter is still a matter of debate. Furthermore, observations reveal that a significant fraction of the intermediate mass and massive stars are born as components of close binaries. We will investigate the effects of binaries on the chemical evolution of Globular Clusters and on the origin of matter with a pristine chemical composition that is needed for the single star AGB scenario to work. We use a population synthesis code that accounts for binary physics in order to estimate the amount and the composition of the matter returned to the interstellar medium of a population of binaries. We demonstrate that the mass lost by a significant population of intermediate mass close binaries in combination with the single star AGB pollution scenario may help to explain the chemical properties of the different populations of stars in Globular Clusters.
}
\maketitle{}

\section{Introduction}

Several Globular Cluster (GC) self-enrichment scenarios have recently been proposed, such as the AGB-scenario (D'Ercole et al. 2010), the winds of fast rotating massive stars scenario (Decressin et al. 2007) and the massive close binary scenario (de Mink et al. 2009). The main challenge in all these is that the enriched material liberated by the first generation of stars needs to be mixed with pristine matter, also present in the vicinity, before the formation of a second generation. The latter then also needs to be sufficiently large, i.e. as least as numerous as the fraction of the first generation that is still observed today. In Vanbeveren et al. (2012) it was shown that intermediate mass close binaries (IMCBs) can help to achieve this. The present paper is a summary of the latter work.

\section{IMCB evolution}

IMCBs are those with an initial primary (the initially most massive star) mass between 3 and 10 M$_{\odot}$, and an initial orbital period between 1 day and 10 years, the latter being the maximum for most binaries to be interacting. There are nowadays many indications that the binary star frequency $f_b$ is very high (close to 100\%) in normal field stars in this mass range. This is not only observed directly, but also inferred indirectly, e.g. by the fact that in theoretical type Ia supernova (SN Ia) studies such as Mennekens et al. (2010), a very high $f_b$ needs to be assumed in order to match the observed rate of such events. The currently observed $f_b$ in (old) GCs, however, seems to be much lower. There are two reasons though why the IMCB frequency in young GCs may have been much higher. Firstly, the IMCB frequency may well be higher than the low mass close binary frequency, with the latter relevant for the GCs currently seen. Secondly, many of the low mass single stars seen in old GCs today may have in fact originally been part of a binary, that has however been destroyed through the Gyrs as a result of dynamical interactions.

Whenever the primary in an IMCB fills its Roche lobe, this will initiate a mass transfer (MT) event. Depending on when this happens, its nature will be different. We distinguish between case A (during core H burning), case B (during shell H burning) and case C (after core He burning). Case A or early case B (named Br, i.e. at a time when the donor's outer layers are still radiative) will result in the expansion of the star being slowed and give rise to a stable stream of matter from donor to gainer, termed Roche lobe overflow (RLOF). For larger orbital periods however, the donor's outer layers are already deeply convective by the time MT (case Bc or C) starts, and thus mass loss will lead to this star expanding even further. It will eventually engulf the other star, resulting in two stellar cores rotating within one common envelope (CE). Some time after this first MT event, when the primary has already become a white dwarf (WD), also the secondary will fill its Roche lobe, initiating a MT event in the opposite direction. Here, also case A, B and C can be distinguished. However, because the gainer is now a WD with a small surface area and the mass ratio in such systems is often very large, it is assumed that this MT will always become unstable and thus result in a CE phase. The only exception are SN Ia progenitors, which are not considered in this study as they do not produce slow winds. During every RLOF or CE phase, the system has a chance of merging if too much angular momentum (AM) is lost. Otherwise, eventually a double WD is obtained, which may still result in a SN Ia but not in a source of slow winds. We will investigate the amount and composition of mass that is lost naturally in a slow way through these RLOF and CE processes.

For this, a population number synthesis (PNS) code including the results of detailed binary evolution is used. This code is extensively described in De Donder \& Vanbeveren (2004). Both the RLOF and CE process are in PNS codes subject to a number of parameterizations. Not only the function and values of these parameters will be discussed, but later also their influence on the results of this study. In the case of stable RLOF, a first parameter is the MT efficiency $\beta$, the fraction of matter lost by the donor that is actually accreted by the gainer. If $\beta<1$, mass is lost into the interstellar medium (ISM) and an assumption needs to be made about how much AM this mass carries away from the system. This is done by a parameter $\eta$, proportional to the AM loss caused by a fixed amount of mass loss. Most PNS codes assume that mass is lost with the specific orbital AM of the gainer, but this is only true in the case of a process that removes mass symmetrically with respect to the equatorial plane of this star. As there is limited observational evidence for significant mass loss in such a way in intermediate mass, non-degenerate stars, our standard assumption is that mass is lost into a circumbinary disk after passing through the second Lagrangian point. It thus removes a much larger specific orbital AM, which will lead to more systems merging. When an unstable CE phase is encountered, the rotational energy of the two rotating stellar cores needs to be converted into kinetic energy used to expel the envelope in time, before the system merges. In the formalism by Webbink (1984), this conversion efficiency is proportional to the parameter $\alpha$, with larger values thus reducing the chance of merging.

\begin{SCfigure*}[][]
\includegraphics[width=8.6cm]{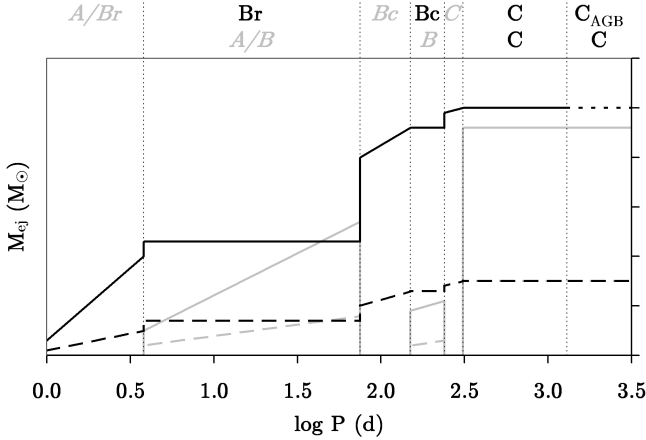}
\caption{\footnotesize
Different types of MT encountered as a function of initial orbital period for a 6.0+5.4 M$_{\odot}$ binary and a non-conservative assumption ($\beta=0.5$). The first line at the top indicates the nature of the first MT phase, the second line (if any) that of the second MT phase. Those in gray italics merge during the particular MT episode. In the plot itself, black lines show the amount of mass loss during the first MT phase, gray lines during the second. This is done for the total mass loss (solid), He mass loss (dashed) and TP-AGB enriched mass loss (dotted).
}
\end{SCfigure*}

It is obvious from the non-conservative example in Fig. 1 that the initial orbital period critically determines the types of MT phases, and thus the amount of lost mass. To illustrate, for the range involving a stable first MT phase (log $P < 1.9$), some mass is lost during this first phase, and some during the second. If the conservative assumption were used instead, obviously no mass would be lost during the first MT phase, but also not during the second, as the gainer would then become sufficiently massive to explode as a SN II. With a slightly lower mass companion (3.6 M$_{\odot}$) and thus mass ratio, the total amount of mass loss is similar in the conservative and non-conservative assumption. While in the latter case it is divided between the first and second MT phase, in the former it is concentrated during the second MT phase, as the secondary becomes much more massive there.

\section{Results}

\begin{SCfigure*}[][]
\includegraphics[width=8.6cm]{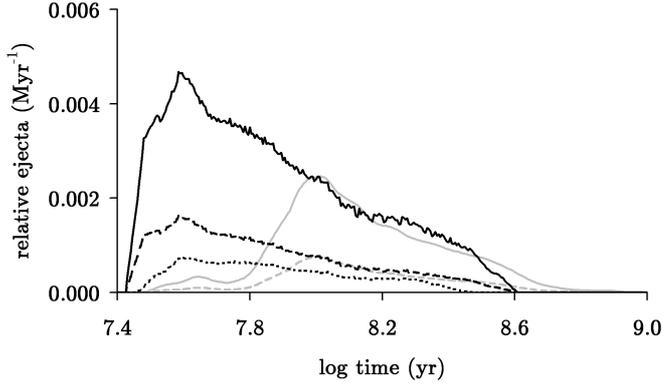}
\caption{\footnotesize
Amounts of mass loss for an entire population of IMCBs under the conservative assumption ($\beta=1$), as a function of time after starburst. The figure shows the total mass loss (solid), He mass loss (dashed) and TP-AGB affected mass loss (dotted), again separated for first (black) and second (gray) MT phase.
}
\end{SCfigure*}

Figure 2 shows that mass loss caused by RLOF and CE processes terminates within 1 Gyr, confirming the previous assertion that this is before dynamical interactions have the chance to destroy many binaries. It is obvious that the two MT phases show two peaks which are population-wise still distinct in time. The general shape of the distribution is not affected when the CE-efficiency parameter $\alpha$ is set to 0.1 instead of 1.0, a value that according to recent observations is more realistic. If one considers the extremely non-conservative case where the gainer does not accrete any material ($\beta=0$), then the mass loss is obviously concentrated at the time of the first peak, and the second peak disappears.

\begin{table*}
\centering
\begin{tabular}{c c c c c c c c c c c}
\hline
Name & $\beta$ & $\eta$ & $\alpha$ & $\Delta M$ & $\frac{\Delta M_{pris}}{\Delta M}$ & $\frac{\Delta M_{pris+He}}{\Delta M}$ & $\frac{\Delta M_{EAGB}}{\Delta M}$ & $\frac{\Delta M_{TPAGB}}{\Delta M}$ & $Y$ & Mergers \\
\hline
conservative & 1.0 & 2.3 & 1.0 & 40\% & 23\% & 22\% & 47\% & 8\% & 0.30 & 76\% \\
low CE-efficiency & 1.0 & 2.3 & 0.1 & 20\% & 14\% & 11\% & 62\% & 14\% & 0.31 & 83\% \\
non-conservative & 0.5 & 2.3 & 1.0 & 34\% & 43\% & 7\% & 40\% & 9\% & 0.28 & 86\% \\
low AM-loss & 0.5 & 0.038* & 1.0 & 53\% & 35\% & 18\% & 41\% & 6\% & 0.29 & 70\% \\
\hline
\end{tabular}
\caption{Results obtained with the PNS code for a population of 100\% IMCBs. See text for definition of symbols. Asterisk denotes weighted average.}
\end{table*}

Table 1 shows the obtained results applied to the context relevant for GCs, for four different sets of parameters: the conservative assumption, the same but with low CE-efficiency, the non-conservative assumption, and again the same but with the specific gainer orbital AM loss, resulting in an (on average) much smaller value of the AM loss parameter $\eta$. The table shows the fraction $\Delta M$ of the initial mass contained in IMCBs that is returned to the ISM in a slow way through RLOF and CE processes. This fraction is between 20 and 50\%. A significant part $\Delta M_{pris}$ of this mass is still pristine, i.e. has not been affected by any nuclear reaction in these stars. A second fraction $\Delta M_{pris+He}$ is enriched only in He, while a third fraction $\Delta M_{EAGB}$ is also enriched by early AGB processes. Only a minor fraction $\Delta M_{TPAGB}$ of about 10\% has been enriched in TP-AGB elements, i.e. those produced by Hot Bottom Burning. The He abundance after the first generation is typically $\sim 0.3$, having started from $Y=0.24$. Extrapolating, after a second generation $Y \sim 0.36$ will be obtained, which is in line with what is needed observationally. It is important to note that of the matter that is not ejected during RLOF and CE processes, the majority will eventually still be returned to the ISM. Only a minor fraction will remain locked in forever as remnant masses and an even smaller fraction will be ejected by SNe (and thus not taken into account here). The reason is that the absolute majority of IMCBs will at some point in their evolution merge (as is also indicated in the table). From this point on, no more RLOF or CE can occur, but matter can still be ejected in normal single star processes (although not with standard single star abundances). Some of these mergers consist of a WD merging with a non-degenerate star, which may lead to very interesting events.

\section{Conclusions}

Intermediate mass close binaries can eject from 20 up to 50\% of their own initial mass in a slow way through naturally occurring Roche lobe overflow and common envelope processes. A large part (20-40\%) of this mass is still pristine, a second part is enriched only in He and CNO-elements, while only a third fraction of about 10\% shows the TP-AGB enrichment typical for the second generation of stars in Globular Clusters. This way, intermediate mass close binaries are not only able to provide a source for the He and TP-AGB enriched material, but also for the pristine matter with which this enriched material needs to be mixed before the formation of a second generation of stars.

\bibliographystyle{aa}

\end{document}